\documentclass[10pt,preprint]{aastex}

\shorttitle{Disrupted Torus Around $\eta$ Car}
\shortauthors{Smith et al.}

\begin{document}

\title{A DISRUPTED CIRCUMSTELLAR TORUS INSIDE ETA CARINAE'S HOMUNCULUS NEBULA\altaffilmark{1}}

\author{Nathan Smith and Robert D. Gehrz}
%% EDITOR - Please do not change affiliation to ``Department of Astronomy...''
\affil{Astronomy Department, University of Minnesota, 116 Church
St. S.E., Minneapolis, MN 55455}
% \email{nathans@astro.umn.edu}

\author{Phillip M. Hinz, William F. Hoffmann, Eric E. Mamajek, and Michael R. Meyer}
\affil{University of Arizona, Steward Observatory, Tucson, AZ 85721}

\and

\author{Joseph L. Hora}
\affil{Harvard-Smithsonian Center for Astrophysics, 60 Garden
Street, MS-65, Cambridge, MA 02138}

\altaffiltext{1}{Based on observations made at the Baade Telescope of
the Magellan Observatory, a joint facility of The Carnegie
Observatories, Harvard University, Massachusetts Institute of
Technology, University of Arizona, and University of Michigan.  }

\begin{abstract}

We present thermal infrared images of the bipolar nebula surrounding
$\eta$ Carinae at six wavelengths from 4.8 to 24.5 $\micron$. These
were obtained with the MIRAC3 camera system at the Magellan
Observatory.  Our images reveal new intricate structure in the bright
core of the nebula, allowing us to re-evaluate interpretations of
morphology seen in images with lower resolution.  Complex structures
in the core might not arise from a pair of overlapping rings or a cool
(110 K) and very massive dust torus, as has been suggested recently.
Instead, it seems more likely that the arcs and compact knots comprise
a warm ($\sim$350 K) disrupted torus at the intersection of the larger
polar lobes.  Some of the arcs appear to break out of the inner core
region, and may be associated with equatorial features seen in optical
images.  The torus could have been disrupted by a post-eruption
stellar wind, or by ejecta from the Great Eruption itself if the torus
existed before that event.  Kinematic data are required to rule out
either possibility.

\end{abstract}

\keywords{ circumstellar matter --- ISM: individual (Homunculus
Nebula) ---stars: individual ($\eta$ Carinae) }

\section{INTRODUCTION}

An elegant expanding bipolar nebula known as the Homunculus surrounds
the persistently peculiar star $\eta$ Carinae.  The nebula contains
several $M_{\odot}$ of material, most of which was ejected during a
major eruption about 160 years ago (see Davidson \& Humphreys 1997).
{\it Hubble Space Telescope (HST)} images at optical wavelengths
(Morse et al.\ 1998) show interesting detailed structure in the polar
lobes, as well as a ragged equatorial debris disk.  Dust in the
Homunculus absorbs most of the star's UV and optical flux, and then
re-emits roughly $4 \ \times \ 10^6 \ L_{\odot}$ at infrared (IR)
wavelengths, making $\eta$ Car one of the brightest sources in the sky
at 10 $\micron$ (Westphal \& Neugebauer 1969), despite its distance of
2.3 kpc (Davidson \& Humphreys 1997).  IR radiation is an especially
useful way to study $\eta$ Car because it reveals structures {\it
inside} the Homunculus that are mostly obscured at shorter
wavelengths.  Numerous investigations at IR wavelengths have shown a
bright, elongated, and multiple-peaked core a few arcseconds across
(e.g., Hyland et al.\ 1979; Mitchell et al.\ 1983, Hackwell et al.\
1986; C.H.\ Smith et al.\ 1995; Rigaut \& Gehring 1995; Smith et al.\
1998; Polomski et al.\ 1999; Smith \& Gehrz 2000).  These features are
usually interpreted as an inclined and limb-brightened circumstellar
torus or disk.  The sharpest mid-IR image of $\eta$ Car so far was
presented by C.H.\ Smith et al.\ (1995), who used a special processing
technique to produce a remarkable 12.5 $\micron$ image showing the
detailed spatial structure of the core, consisting of several knots
and loop structures.  Smith et al.\ (1998) and Polomski et al.\ (1999)
discussed multi-wavelength IR array images of $\eta$ Car, and
summarized this object's unique thermal-IR emission properties.

A controversey has recently arisen regarding the nature of the bright
structures in the core of the Homunculus.  Several previous observers
had described the torus and showed that it had a color temperature
higher than 250 K at wavelengths near 10 $\micron$ (e.g., Hackwell et
al.\ 1986; C.H.\ Smith et al.\ 1995; Smith et al.\ 1998; Polomski et
al.\ 1999).  However, Morris et al.\ (1999) recently claimed to have
discovered this feature, and their interpretation of a large-aperture
2 to 200 $\micron$ spectrum obtained with the {\it Infrared Space
Observatory (ISO)} led them to associate the torus with much cooler
dust at 110 K.  They proposed that this cool torus contained $15 \
M_{\odot}$ of material, and was formed when a companion star in a
close binary system stripped the normal composition envelope off the
primary star before the Great Eruption.  They further suggested that
the ejection of this massive torus caused the Great Eruption (by
increasing the star's $L / M$ ratio) and was directly responsible for
the bipolar shape of the Homunculus Nebula.  However, Davidson \&
Smith (2000) showed that the hypothesized compact torus inside the
Homunculus would be too small to radiate the required luminosity with
a brightness temperature of only 110 K. Hony et al.\ (2001) proposed
instead that the inner dust was warmer, as already shown by earlier
studies, but they interpreted structure seen in their images as a pair
of overlapping rings with a geometry analogous to those around SN
1987a (e.g., Burrows et al.\ 1995).  These hypothetical rings have a
different polar axis than the Homunculus --- so Hony et al.\ (2001)
made the extraordinary claim that an interaction between three bodies
had caused the orbital axis to change orientation by more than one
radian during or after the Great Eruption in the mid-19th century when
the Homunculus was ejected.  This interpretation seems unlikely for
several reasons, and here we present new IR data showing that
interpreting the IR features inside the Homunculus as a pair of rings
is less straightforward than Hony et al.\ suggest.  Instead these
features could comprise an equatorial disk of material as previously
proposed, which appears significantly disrupted when observed at high
spatial resolution.

In \S 2 we present our new thermal-IR images, and in \S 3 we discuss
the IR structures observed in these high-resolution images, as well as
their consequences for previous interpretations of $\eta$ Car's IR
radiation.  This will be a brief and descriptive analysis of the IR
structures in the core of the nebula; a more thorough quantitative
analysis of these data will follow in a later paper.

\section{OBSERVATIONS}

Thermal-IR images of $\eta$ Carinae were obtained on 2001 August 8
with the MIRAC3/BLINC instrument mounted on the Baade 6.5m telescope
of the new Magellan Observatory.  The images presented here were among
the first thermal-IR data obtained with the Magellan telescopes.
MIRAC3 is a mid-IR array camera built for ground-based astronomical
imaging at Steward Observatory, University of Arizona, and the
Harvard-Smithsonian Center for Astrophysics (Hoffmann et al.\ 1998).
It utilizes a Rockwell HF-16 128$\times$128 hybrid BIB array, and is
an upgrade from the original MIRAC system (Hoffmann et al.\ 1994).
Standard chop-nod sets of images were obtained using six narrow
filters at the wavelengths listed in Table 1, which summarizes other
observational parameters.

The new images have better spatial resolution than any previous images
of $\eta$ Car at these wavelengths, and reveal new and important
structures.  However, $\eta$ Car was observed primarily as a test of
MIRAC3's image quality on the new Baade 6.5m telescope, and so proper
observations for flux calibration before and after each observation
were not performed.  The standard star $\gamma$ Cru was observed just
after $\eta$ Car on the same date at 4.8 and 18.0 $\micron$, but the
only acceptable calibration observations at remaining wavelengths were
obtained a few nights later on 2001 August 14, using HD~169916 as the
standard star.  The 18.0 $\micron$ filter was used on both occasions,
and the resulting calibration for the two observations agreed to
within 10\%.  Despite these difficulties, the resulting {\it relative}
fluxes between various filters agreed with those expected from the
{\it ISO} spectrum of $\eta$ Car published by Morris et al.\ (1999),
so the data are useful for investigating spatial variation in color
temperature and other emission properties of the dust.  The relative
photometric accuracy is $\sim$10\%, mainly due to uncertainty in the
calibration system.

After sky subtraction, airmass correction, and gain correction,
individual frames were combined with pixels subdivided by a factor of
2, for a final pixel scale of 0$\farcs$061.  Numerous frames were
obtained for each filter, which allowed us to reject those with
unsatisfactory seeing or focus.  False-color representations of the
final images in each of the six MIRAC3 filters are shown in Figure 1.

Only the 10.3 $\micron$ filter is severely affected by emission from
silicates, so we can use the other filters to estimate the grain color
temperature and emission opacity at each position in the nebula, after
applying small corrections for wavelength-dependent spatial resolution
(see Table 1).  Figure 2$a$ shows a map of the 12.5 to 18.0 $\micron$
color temperature distribution, and the corresponding 18 $\micron$
emission optical depth is shown in Figure 2$b$.  These maps were
constructed using methods that are standard for mid-IR array imaging
(for instance, Polomski et al.\ 1999 explain how to derive $T_c$ and
$\tau$ maps).  We assumed that the grains have an emissivity
proportional to $\lambda^{-1}$, appropriate for amorphous silicates
with $a \ \approx 1~\micron$.

\section{MORPHOLOGY OF THE BRIGHT INFRARED CORE}

The diffraction limit of the Magellan~{\sc i} telescope's 6.5m primary
mirror has allowed us to produce high-quality IR images of the
Homunculus Nebula without using extensive image enhancement routines.
The images in Figure 1 give a much sharper picture of structures near
the star seen previously in the 10 to 20 $\micron$ continuum (C.H.\
Smith et al.\ 1995; Hony et al.\ 2001), the near-IR continuum (Rigaut
\& Gehring 1995; Smith \& Gehrz 2000), and near-IR Br$\gamma$ emission
(Smith et al.\ 1998).

Several compact knots reside in the core; the best example is the
bright spot $\sim$1$\arcsec$ SE of the star in the 8.8 $\micron$ image
in Figure 1$b$.  These knots could be part of an equatorial torus, but
if so, they are not equidistant from the star (the hypothetical torus
has some azimuthal asymmetry).  Figure 2 suggests that, in general,
the bright spots are found at the warm heads of protruding regions of
relatively high density.  The brightest knot, located less than
0$\farcs$5 to the NW of the star, coincides with compact {\it
equatorial} emission-line blobs seen in high-resolution optical
studies (Hofmann \& Weigelt 1988; Davidson et al.\ 1995).

Thin filaments that form arcs or loops appear to connect adjacent IR
knots in the core.  In some directions these features appear to break
through the constraining dust torus (extended features are seen best
at 18 and 24.5 $\micron$), and coincide with prominent equatorial
features in optical {\it HST} images (see Morse et al. 1998).
However, the relationship between the IR (thermal dust emission) and
optical (scattered light) structures is unclear.  These arcs resemble
some remarkable features seen recently in {\it HST} images of the red
supergiant VY CMa (Smith et al.\ 2001a). The prominent arc 2$\arcsec$
SE of the star clearly connects with the limb-brightened rear wall of
the receeding polar lobe in the 18 and 24.5 $\micron$ images.  In
fact, the dust optical-depth map in Figure 2$b$ suggests that the
observed structures in the core are part of a torus that constitutes
the intersection of the two polar lobes, as depicted schematically in
Figure 3.  Limb-brightening of the walls of both polar lobes also
contributes to the observed structure; Figure 2$b$ shows a morphology
expected for an overlapping pair of mostly hollow osculating spheres.

Hony et al.\ (2001) envisioned a pair of overlapping rings with a
symmetry axis very different from the Homunculus as the physical
explanation for features that resembled rings or loops in their 8 to
20 $\micron$ images.  Our 24.5 $\micron$ image, which has slightly
improved spatial resolution compared to their images, shows one
loop-like feature that resembles the brighter of the two rings
described by Hony et al.  However, our images at shorter wavelengths
show that with higher spatial resolution, the illusory loop breaks up
into a set of compact knots and incomplete arcs.  Thus, our images
show that interpreting the complex IR morphology of the inner
Homunculus as a pair of smooth rings is less definite than Hony et
al.\ suggest.  The geometry they propose cannot be reconciled with the
fact that the brightest section of the purported rings is coincident
with the optically identified ``Weigelt blobs'', whose kinematics
indicate that they are indeed equatorial (Davidson et al.\ 1995).
Furthermore, recent results from optical {\it HST} spectroscopy of
$\eta$ Car contradict Hony et al.'s claim that the orbital axis of the
central system has precessed since the Great Eruption in 1843:
Kinematics of ionized gas inside the Homunculus indicate that some
material ejected around 1890 has the same bipolar axis as the
Homunculus (Ishibashi et al.\ 2001), and reflected spectra of the
central star show that its current polar axis is still aligned with
the Homunculus (Smith et al.\ 2001b).  Since the putative smooth
double-ring structure may be an artifact of insufficient spatial
resolution, Hony et al.'s suggestion that tidal interactions after the
Great Eruption changed the polar axis of the hypothetical binary
system remains unsubstantiated.

Another possible interpretation of the bright IR structures in the
core of the Homunculus invokes a toroidal distribution of dust (at
$\sim$3000 AU from the star), although the torus must have suffered
significant disruption.  Suppose an initial torus or disk existed
around the star, and that stellar wind or ejecta then plowed into this
material, as often invoked in hydrodynamic simulations of bipolar
nebulae (Frank et al.\ 1995; Dwarkadas \& Balick 1998).  Dense knots
in that torus would not be accelerated as quickly as material between
them, forming the protruding dense knots and more extended arcs seen
in our images (note that the knots are all closer to the star than
adjacent arcs; see also Figure 3).  The dense knots could be
pre-existing density enhancements, or perhaps could result from
Rayleigh-Taylor instabilities.  In either scenario, the knots would be
analogous to bright ends of dust pillars often seen in star forming
regions, but on a much smaller size scale around $\eta$ Car.  So far,
2-D numerical simulations do not predict such structures, but we
expect that they may be seen in fully 3-D simulations.  Alternatively,
one can imagine more complicated mass-ejections with preferred
directions that disrupt the torus.  We cannot yet rule out this
possibility.  Based on our images, we cannot infer the age or exact
origin of the equatorial structures; kinematic information is needed.
Comparing the distribution of dust emission and ionized gas (for which
kinematics are available) may help clarify the inner geometry of the
Homunculus.

Our analysis of the new IR images presented here suggests that the
bright structures in the core of the nebula are indeed part of a
circumstellar equatorial torus or disk, as proposed previously by
several authors.  This could have important implications for shaping
the bipolar Homunculus and its ragged equatorial debris disk.  The
nature and location of the excess far-IR flux seen in the {\it ISO}
spectrum (Morris et al.\ 1999) remains unanswered; resolving this
dilemma is important because it means that several $M_{\odot}$ of
material in the Homunculus may be unaccounted for, which would have
important implications for the energy budget during $\eta$ Car's Great
Eruption.  The temperature map in Figure 2$a$ suggests that some of
the missing cool dust may actually reside in the polar lobes; we will
address this separate problem in a future publication with a more
thorough quantitative analysis.

\scriptsize
\acknowledgements

N.S. is grateful for the support of a NASA/GSRP fellowship from
Goddard Space Flight Center.  R.D.G. is supported by the NASA SIRTF
program and the U.S. Air Force. MIRAC is supported by National Science
Foundation grant AST 96-18850, and BLINC was developed under a
NASA/JPL grant for TPF.  M.R.M. and E.E.M. gratefully acknowledge
support through NASA contract 1224768 administered by JPL.

%%% TABLE 1
\begin{table}
\caption{MIRAC3 Observations (2001 Aug 8)}
\begin{tabular}{lcccl} \tableline\tableline
$\lambda$	&$\Delta\lambda$	&exp.\tablenotemark{a}	&FWHM	&standard star	\\

($\micron$)	&($\micron$)	&(sec)	&(arcsec)	&		\\	\tableline

4.8		&0.77	&646	&0.31	&$\gamma$ Cru	\\
8.8		&0.88	&60	&0.35	& HD 169916	\\
10.3		&1.03	&46	&0.40	& HD 169916	\\
12.5		&1.16	&46	&0.36	& HD 169916	\\
18.0		&1.80	&78	&0.42	&$\gamma$ Cru, HD 169916 \\
24.5		&1.91	&158	&0.61	& HD 169916	\\
\tableline
\end{tabular}
\tablenotetext{a}{Total on-source exposure time.}
\end{table}

%%%%%%%%%% FIGURE 1
\begin{figure}
\plotone{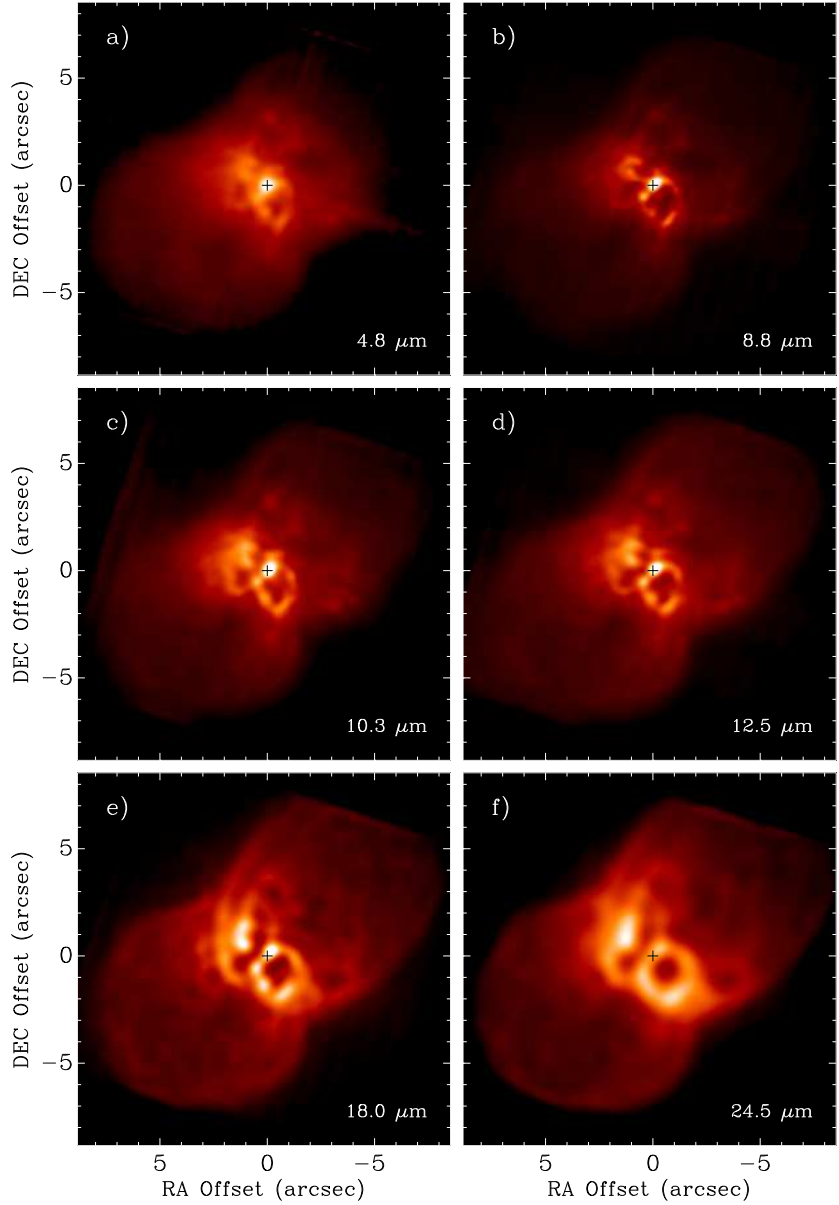}
\caption{MIRAC3/Magellan images of the Homunculus at the indicated
wavelengths, shown in false-color using the square root of the
observed intensity to accommodate the large dynamic range in the
images.  The faint streaks at the left edge of panel (c) and the
straight termination of the upper-right end of the Homunculus in
panels (e) and (f) are artifacts due to the edge of the detector
array.  In each panel, the black plus-sign marks the position of the
central star.}
\end{figure}

\begin{figure}
\epsscale{0.45}
\plotone{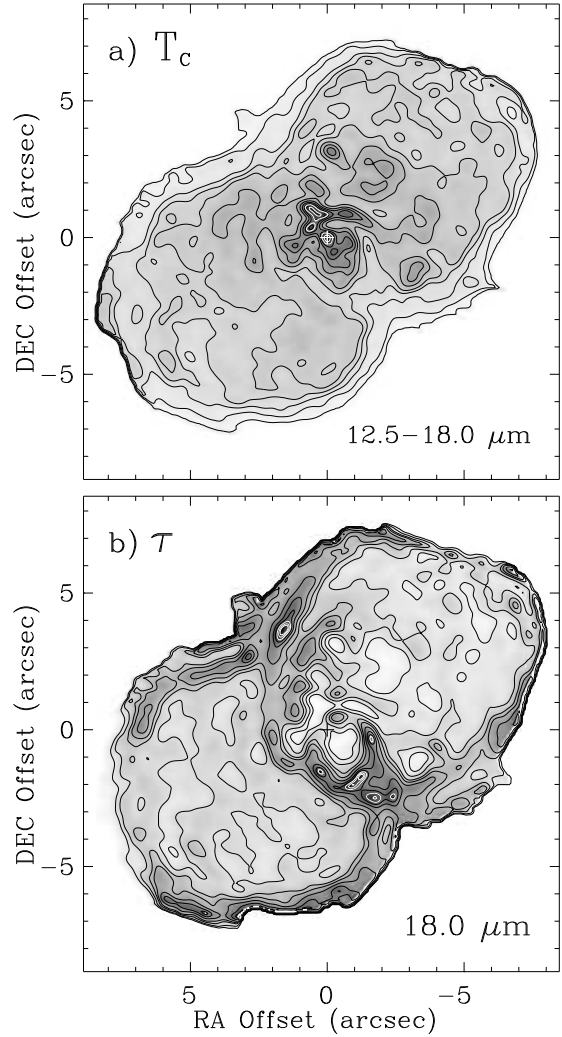}
\caption{(a) Map of the observed dust color temperature
resulting from the 12.5 to 18.0 $\micron$ flux ratio.  Contours are
drawn at 120, 140, 160, 180, 200, 240, 280, 320, 360, 400, 450, 550,
and 650 K.  (b) Corresponding map of emitting optical depth ($\tau$)
of dust in the Homunculus at a wavelength of 18 $\micron$.  Contours
are drawn at values of 0.05, 0.1, 0.2, 0.3, 0.5, 0.7, 1.0, 1.3, 1.6,
and 2.0.  Contour levels for $\tau \ > \ 1$ are white.  The plus-sign
marks the position of the central star in both panels.}
\end{figure}

\begin{figure}
\epsscale{0.55}
\plotone{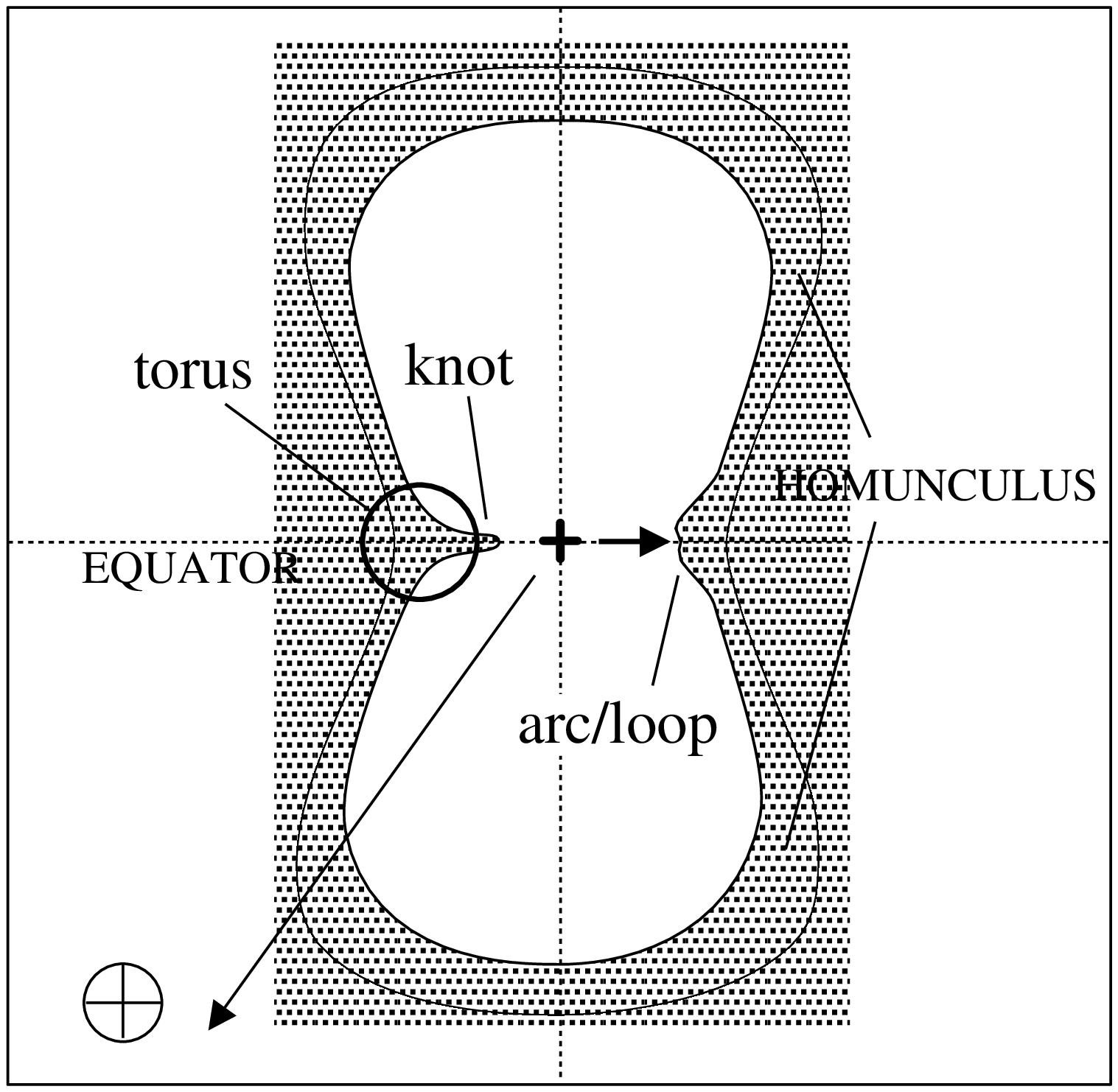}
\caption{Cartoon showing a schematic cross-section of the Homunculus.
The region where the two polar lobes meet at the equatorial plane
forms a torus.  The figure depicts knots that protrude into the nebula
in the equatorial plane, as well as arcs that are swept back by
stellar wind (the dark arrow).  See \S 3.}
\end{figure}


\begin{references}

\reference{}Burrows, C.J., et al.\ 1995, ApJ, 452, 680

\reference{}Davidson, K., \& Humphreys, R.M.\ 1997, ARAA, 35, 1

\reference{}Davidson, K., \& Smith, N.\ 2000, Nature, 405, 532

\reference{}Davidson, K., Ebbets, D., Weigelt, G., Humphreys, R.M.,
Hajian, A., Walborn, N.R., \& Rosa, M.\ 1995, AJ, 109, 1784

\reference{}Dwarkadas, V.V., \& Balick, B.\ 1998, AJ, 116, 829

\reference{}Frank, A., Balick, B., \& Davidson, K.\ 1995, ApJ, 441, L77

\reference{}Hackwell, J.A., Gehrz, R.D., \& Grasdalen, G.L.\ 1986, ApJ, 311, 380

\reference{}Hoffmann, W.F., Fazio, G.G., Shivanandan, K., Hora, J.L.,
\& Deutsch, L.K.\ 1994, Infrared Phys. Tech., 35, 175

\reference{}Hoffmann, W.F., Hora, J.L., Fazio, G.G., Deutsch, L.K., \&
Dayal, A.\ 1998, in Infrared Astronomical Instrumentation, ed.\ A.M.\
Fowler, Proc. SPIE 3354, 647

\reference{}Hofmann, K.H., \& Weigelt, G.\ 1988, A\&A, 203, L21

\reference{}Hony, S., Dominik, C., \& Waters, L.B.F.M., et al.\ 2001,
A\&A, 377, L1

\reference{}Hyland, A.R., Robinson, G., Mitchell, R.M., Thomas, J.A.,
\& Becklin, E.E.\ 1979, ApJ, 233, 145

\reference{}Ishibashi, K., Gull, T.R., \& Davidson, K.\ 2001, in ASP
Conf.\ Ser.\ 242, Eta Carinae and Other Mysterious Stars, Ed. T.R.\
Gull, S.\ Johansson, \& K.\ Davidson (San Francisco: ASP), 71

\reference{}Mitchell, R.M., Robinson, G., Hyland, A.R., \& Jones,
T.J.\ 1983, ApJ, 271, 133

\reference{}Morris, P.W., Waters, L.B.F.M., Barlow, M.J., et al.\
1999, Nature, 402, 502

\reference{}Morse, J.A., Davidson, K., Bally, J., Ebbets, D., Balick,
B., \& Frank, A.\ 1998, AJ, 116, 2443

\reference{}Polomski, E., Telesco, C.M., Pina, R.K., \& Fisher, S.F.\
1999, AJ, 118, 2369

\reference{}Rigaut, F., \& Gehring, G.\ 1995, RevMexAA, Ser.\ Conf., 2, 27

\reference{}Smith, C.H., et al.\ 1995, MNRAS, 273, 354

\reference{}Smith, N., \& Gehrz, R.D.\ 2000, ApJ, 529, L99

\reference{}Smith, N., Gehrz, R.D., \& Krautter, J.\ 1998, AJ, 116, 1332

\reference{}Smith, N., Humphreys, R.M., Davidson, K., Gehrz, R.D.,
Schuster, M.T., \& Krautter, J.\ 2001a, AJ, 121, 1111

\reference{}Smith, N., Davidson, K., Gull, T.R., \& Ishibashi, K.\
2001b, in ASP Conf.\ Ser.\ 242, Eta Carinae and Other Mysterious Stars,
Ed. T.R.\ Gull, S.\ Johansson, \& K.\ Davidson (San Francisco:
ASP), 117

\reference{}Westphal, J.A., \& Neugebauer, G.\ 1969, ApJ, 156, L45


\end{references}
\end{document}